\documentstyle[12pt,a4,psfig]{article}
\def\eqnum#1{\eqno (#1)}
\title { \bf On production $e^{+} e^{-}$-pairs by a high energy photon in
collision with photons of a laser wave \footnote{Published in: Proceedings
of the V Annual seminar "Nonlinear phenomena in complex system".
February 1996, Minsk, Belarus. Advances in Synergetics. 1997, Vol.8, pp.60-63,
Minsk. Edited by V.Kuvshinov \& G.Krylov.} }

\author {\large Mikhail Galynskii \footnote{Email: galynski@dragon.bas-net.by}
 ~and  Sergei Sikach \footnote{Email: sikach@dragon.bas-net.by}\\
\it Stepanov Institute of Physics Academy of Sciences of Belarus \\
\it Fr.Scorina av.70, 220072, Minsk, Belarus }
\date{}
\begin{document}
\maketitle

\begin{abstract}
We calculate the number of $e^{+}e^{-}$-pairs produced by
a Compton high energy  photon in turn formed in the process
$ e + n \gamma _{0}  \rightarrow  e + \gamma  $
at simultaneous  collision  with several photons $\gamma _{0}$ of the same
laser beam $\gamma + s \gamma _{0} \rightarrow e^{+} + e^{-}$.
The case when polarization states of initial electron beam and laser
photons are helicity is considered.
It is shown that taking into account  nonlinear effects  in
the Compton backscattering leads to decreasing the threshold
of production of $e^{+}e^{-}$-pairs and increasing
their number.
The consideration is based on the direct
calculation of matrix elements in the formalism of diagonal spin
basis.
\end{abstract}

In [1,2] production of colliding $\gamma e$-
and  $ \gamma \gamma$-beams in singlepass
accelerators have been suggested. So, intense beams of hard
$\gamma $-quantums are to be produced in the Compton backward
scattering of photons of power laser pulse that is  focused on a
beam of ultrarelativis\-tic electrons.  At sufficient power of the
laser pulse, the process of emission of  a high energy photon can be
realized through absorption of several laser photons
$$
e^{-} \; + \; n \; \gamma _{0} \; \rightarrow  \;
e^{-} \; + \; \gamma \; , \qquad  n\; \geq \;  1
\eqnum{1}
$$

\noindent   the latter leads to the widening of
spectrums  of hard  $\gamma $-quantums [3,4]. In [5], it was shown
that the hard photon, produced in the reaction (1),
can create $e^{+} e^{-}$-pairs at
collisions with photons of the same laser beam
$$
\gamma \; + \;s\;  \gamma _{0} \; \rightarrow \;
  e^{+}\; + \; e^{-}    \; ,\qquad s\; \geq 1 \; .
\eqnum{2}
$$

\noindent The threshold of this reaction at $s = 1$ is very high.
Minimal value of the Compton photon energy for the process (2)
(at the use of a neodymium laser with  $\omega _{0} =
1.17$ eV) is equal to   $\omega  = m^{2} / \omega _{0}  = 223 $ GeV.
In reality,   $e^{+} e^{-}$-pairs will be created in a
great  number even at noticeably smaller energies due to
simultaneous collisions of a hard
photon $\gamma $ with several laser photons.

Observation of the process (2) represents doubtless interest for
verification of the quantum electrodynamics at a new range of
parameters.  In the same time, it provides  substantial background
for $\gamma  e$-  and  $\gamma  \gamma $-collisions [2].

The processes (1) and (2) represent nonlinear interactions
of  electrons and photons, respectively,  with  the field of
electromagnetic wave. It is easy to see that the
influence of those nonlinear effects  in the process (1) generates,
in the process (2),  significant decreasing of the threshold
of  the production of
$e^{+} e^{-} $-pairs and  the increasing  of their number.

A maximal energy of the Compton  photon $\gamma $ obtained as a result
of absorption, by a single electron with energy $E$,
of $n$ laser photons (with the energy  $\omega _{0}$)
is equal to
$$
\omega _{n} = \; { n \kappa  \over 1 + n \kappa } E \;
,\qquad \kappa  = { 4 \omega _{0} E \over m^{2}}   \; .
\eqnum{3}
$$

The energy threshold of the photon  $\gamma $  (for
the process (2)) is determined from the relation $$ ( k\; + \;s\;
k_{0} )^{2} = 4 \;m^{2}   \; ,
\eqnum{4}
$$

\noindent where  $k$ and  $k_{0}$ are  $4$-momentums of
the photons $\gamma $ and $\gamma _{0}, \; m$ is the electron rest mass.

\noindent Corresponding threshold values of energy $E_{ns}$   of electrons in
a beam (for producing  $e^{+} e^{-}$-pair through absorption of  $n$
photons and  subsequent collision of the hard photon with  $s$ laser
photons) are determined with the aid of (3), (4) and they are equal to
$$
E_{ns}  =
{m^{2}  \over 2 \; \omega _{0}\; s} \; (1 \;+\;  ( 1 + s / n )^{1/2} )\; .
\eqnum{5}
$$

\noindent With the use of (5) we  have calculated the values
$E_{1s}$  and $E_{2s}$   at  $ 1 \leq  s \leq  6$:
$$
Table \;  1.  \qquad
\left. \begin{array}{cccccccc}
 s      &     & 1    &  2   & 3   &  4  &  5  &  6   \\
 E_{1s} & GeV & 269  &  153 & 112 &  90 &  77 & 68   \\
 E_{2s} & GeV & 248  &  135 & 96  &  76 &  64 & 56
\end{array} \right.
$$

\noindent These results evidently say that the widening of
spectrums of hard $\gamma $-quantums due to the above nonlinear
effects leads to  the decreasing of the threshold of production of
$e^{+} e^{-}$-pairs.

According to our calculation (details of the used method calculation
amplitudes in the diagonal spin basis can be seen in the authors' work [6]),
the differential probability of the process (2), in the field of
circularly  polarized electromagnetic wave, is given by the relations
$$
d W^{(s)} =
{ e^{2} m^{2} \over 4 \pi \omega } \;
\mid  M^{(s) \lambda \lambda'}_{\pm \mu , \mu } \mid ^{2}  \;
\delta ^{4} (s k_{0} + k - q - q' ) \;
{ d^{3} q \; d^{3} q'  \over q_{0} \; q_{0}' }
\eqnum{6}
$$
$$
 M^{(s)\lambda \lambda'}_{\mu ,\mu } =
(-\lambda)^{s}  \left [- \lambda ' \mu n_{1} n_{3}' J_{s} +
{\xi m s \over m^{2}_{*}  u_{s}} \;
(u k n_{0}' \;- \; \lambda' \mu  \epsilon \; \sqrt{u (u-1)} \;
k n_{3}' )
\; J_{s-\lambda \lambda '} \right ]
\eqnum{7}
$$
$$
 M^{(s) \lambda \lambda ' }_{-\mu , \mu } =
- \lambda '(-\lambda)^{s} (n_{1} n_{1}' + \lambda'\mu)
\left [\sqrt{(vv'+ 1)/2} J_{s} +
{\xi m s u \over m^{2}_{*} u_{s} }
 \sqrt{(vv'-1)/2} k n_{1}'J_{s-\lambda \lambda'}\right ]
\eqnum{8}
$$

\noindent where
$$
n_{1} \; n_{3}'  =
- {m^{2}_{*} \; u_{s} \over m^{2} u  \sqrt{2(v v'- 1 ) } } \;
{z \over s \xi} \;  ,  \qquad
k n_{0}' = {2 m^{2}_{*} u_{s} \over m s \sqrt {2(v v' + 1)} } \;  ,
$$
$$
k n' _{3} = - \epsilon  {2 m^{2}_{*} u_{s} \sqrt{ (u-1)/ u }
\over m s \sqrt{2(vv' - 1) } } \; ,  \qquad
n_{1} n'_{1} = \epsilon  \sqrt {{u-1 \over u }}
\sqrt {{vv' + 1 \over vv' -1 }} \; ,
$$
$$
k n' _{1} = - {m^{4}_{*} u^{2}_{s} \over s^{2} m^{3} u
\sqrt{ (vv')^{2} - 1 }} {z \over \xi } \;  , \qquad
\epsilon  = \hbox{ sign } \sqrt{{u_{s} (u-1) \over u (u_{s}-1)}} \; ,
$$
$$
u = {(k k_{0})^{2} \over 4 k_{0} q \; k_{0} q' } \; , \; \;
u_{s}= {s k k_{0} \over 2 m^{2}_{*}} \; , \; \;
z = {2 s \xi \over \sqrt{1 + \xi^{2} }} \sqrt { { u \over u_{s}}
(1- {u \over u_{s}} })\; ,
$$
$$
J_{s-\lambda \lambda '} = {1 \over 2} (1+\lambda \lambda ')\;
 J_{s-1} \; + \; {1 \over 2} (1- \lambda \lambda ') J_{s+1}\; ,
$$
$$
 q = p + {\xi^{2} m^{2} \over 2 k_{0} p } k_{0} \; ,\qquad
q'  = p'  + {\xi^{2} m^{2} \over 2 k_{0} p' } k_{0} \; , \qquad
q^{2} = (q')^{2} = m^{2}_{*} = m^{2} (1 + \xi^{2})\;  ,
$$
$$
s k_{0} + k = q + q' \; ,\qquad
 v v'  - 1 = 2 (u_{s} - 1 + \xi^{2} (u_{s} - u) ) \; .
$$

\noindent Here $k_{0}$ and  $\lambda$  ($k$ and $\lambda '$)  are
4-momentum  and  helicity  of a laser (hard) photon ;
$q$ and  $q'$ are quasi-momentums of the positron and electron,
$J_{s} = J_{s}(z) $ are the Bessel
functions, $\xi^{2}$ is a laser field-strength parameter.

For total probability (per second)  of pairs production in the process (2) we vave:
$$
W={\alpha m^2 \over 4 \; \omega} \sum^{\infty}_{ s>s_0 }\; \int \limits ^{u_{s}}_{1}\;
(F_{0s} + \lambda \lambda' F_{2s} +\mu \lambda G_{0s} + \mu \lambda' G_{2s})\;
{du \over u \sqrt{u(u-1)}}\; ,
\eqnum {9}
$$
$$
F_{0s} = J_{s}^2 + \xi^2 \; (2u-1)\; (-J_{s}^2 +(J_{s-1}^2 + J_{s+1}^2)/2\; )\; ,~~~~~~~~~~~
$$
$$
F_{2s} = \xi^2 \; (2u -1) (2u /u_s -1 ) (J_{s-1}^2 - J_{s+1}^2)/2\; ,~~~~~~~~~~~~~~~~
$$
$$
G_{0s} = \psi_{+}\psi_{-} \; \xi^2 \; u/u_s (u_s -1) \; (J_{s-1}^2 - J_{s+1}^2\; )\; , ~~~~~~~~~~~~~~~~~~~
$$
$$
G_{2s} = \psi_{+}\psi_{-} \; \{ u_s J_s^2 +\xi^2 [u_s J_s^2 +u(u-1)
(J_{s-1}^2 + J_{s+1}^2)\; ] \} \; , ~~
$$
where  $\psi_{\pm} = 1/\sqrt{(pp' \pm m^2)/2m^2}, \; \mu$-- is the value of
positron spin projection in the diagonal spin basis [6] ($\mu=\pm 1),
\alpha $ is the fine-structure constant.

The total number $N_{ e^{+} e^{-} }$ of produced  $e^{+}e^{-}$-pairs
is obtained by summation on energy spectrum of Compton photons [5]:

$$ N_{e^{+}e^{-}} = N_{\gamma }\;
 {\tau \over 4} \; \sum^{\infty}_{s_{0}} \; \int^{\omega _{n}}_{ 0}
 W^{(s)} (\omega , \omega _{0}, \xi) \; {1\over \sigma _{c}(E)}\;
{d\sigma _{c} \over d\omega }\;  d\omega \; ,
$$

\noindent where $N_{\gamma }$ is the total number of hard photons,
$\sigma _{c}(E)$ and $d\sigma _{c}/ d\omega $ are total and
differential cross sections of Compton scattering,
$W^{(s)}(\omega , \omega _{0} , \xi)$  is  the probability of
production of a pair by a hard photon (the
case of a circularly  polarized laser wave is  considered).

Results of numerical calculations for
 $\ln ( N_{e^{+}e^{-}} / N_{e})$ in the diagonal spin basis,
depending on the electron energy $E$,  for different values
of energy $A$ of a laser pulse,  helicity
of laser photons $\lambda$  and initial electrons $\lambda _{e}$,
 are given in Figures 1 and 2. For Nd:glass laser at $\omega_0=1.17$ eV
and electron beam parameters we have choosen [5]: $c\tau=1$ cm,
$\pi r_e^2 = \pi a_{\gamma}^2 =10^{-5}$ cm$^2$, where $c$ - is the speed
of light.

\begin{figure}
\begin{minipage}[t]{8cm}
\psfig{file=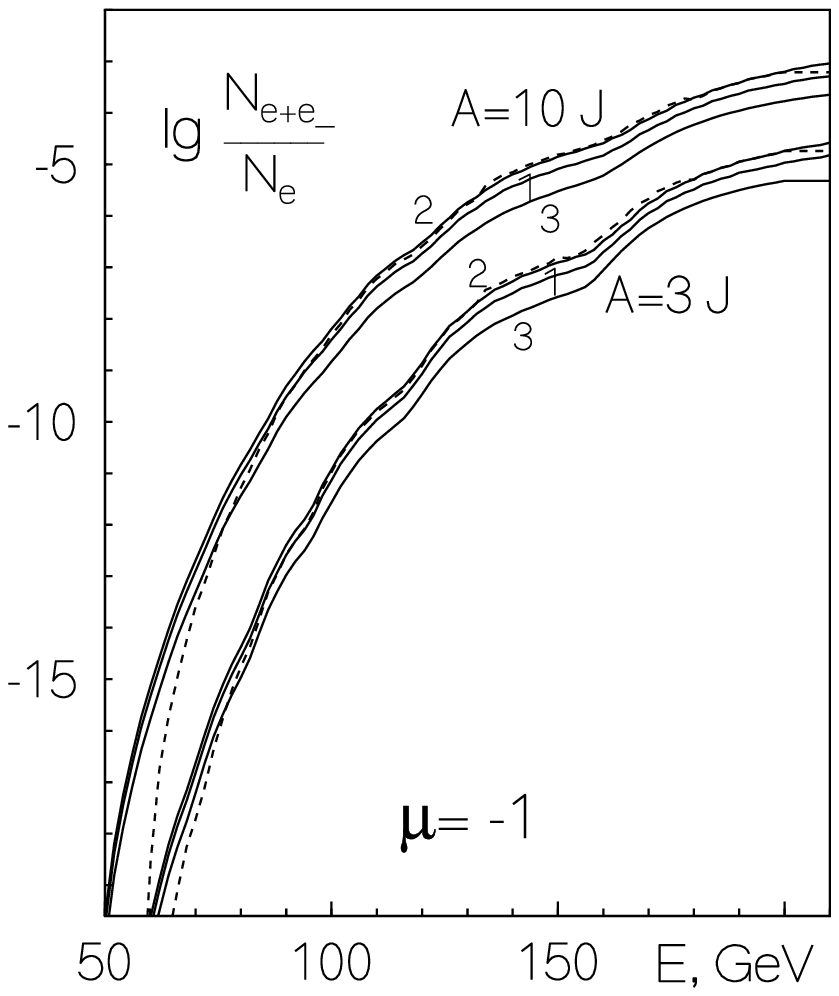,width=8cm}
\end{minipage}
\begin{minipage}[t]{8cm}
\psfig{file=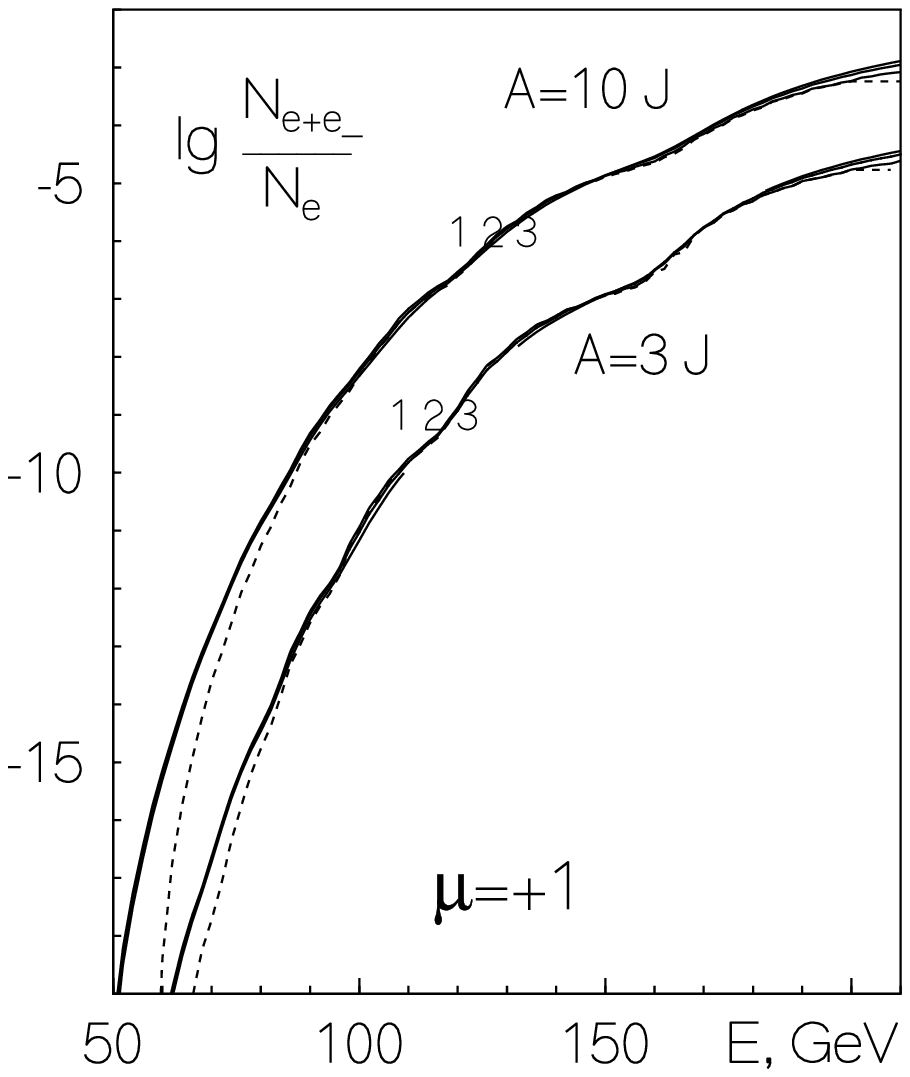,width=8cm}
\end{minipage}

\vspace{-1.5cm}
\hspace{3.5cm} Fig.1 \hspace{7cm} Fig.2

Number of $e^{+} e^{-}$-par production per one electron vs of electron bunch
energy $E$ for laser flash energy $A=3 J$ and $A=10 J$. Left and right figures
corespond to positron spin projection $\mu=-1$ and $\mu=+1$ respectively.
The lines $1,\;2,\;3$ there correspond to the three choices of
helicity in the process (1): $1) \; \lambda \lambda _{e} = 0 \; , \; 2) \;
\lambda \lambda _{e} = -1 ,\; 3) \; \lambda \lambda _{e} = 1$.
The solid lines are related to $n = 2$, the dashed lines -- to $n = 1$.
\end{figure}

 \vspace{3cm}

From the Figures 1, 2 it follows that the nonlinear effect in the
process (1) leads to the significant increasing of the number of
$e^{+}e^{-}$-pairs  at the range of energies of present-day accelerators [7]
 and future high energy linear colliders [8].

\end{document}